\documentclass[aps,prd,twocolumn,showpacs,groupedaddress,preprintnumbers]{revtex4}
\usepackage{graphicx}% Include figure files
\usepackage{dcolumn}   % needed for some tables
\usepackage{bm}        % for math
\usepackage{amssymb}   % for math
\usepackage{color}
\begin{document}
%\setlength{\topmargin}{.1cm}
%\linenumbers
\preprint{hep-ex/0902.3451v4}
%\bibliographystyle{apsrev}

% Use the \preprint command to place your local institutional report
% number in the upper righthand corner of the title page in preprint mode.
% Multiple \preprint commands are allowed.
% Use the 'preprintnumbers' class option to override journal defaults
% to display numbers if necessary
%\preprint{}
%Title of paper
\title{Search Capability for $\eta\to\nu_{e,{\tau}}\bar\nu_{e,{\tau}}$ Decays
in Cubic Kilometer Neutrino Detectors}

% repeat the \author .. \affiliation  etc. as needed
% \email, \thanks, \homepage, \altaffiliation all apply to the current
% author. Explanatory text should go in the []'s, actual e-mail
% address or url should go in the {}'s for \email and \homepage.
% Please use the appropriate macro foreach each type of information

% \affiliation command applies to all authors since the last
% \affiliation command. The \affiliation command should follow the
% other information
% \affiliation can be followed by \email, \homepage, \thanks as well.
%\homepage[]{Your web page}
%\thanks{}
%\altaffiliation{}
%\affiliation{Southern University, Baton Rouge, LA 70813}
\author{A.R. Fazely}\email{ali_fazely@subr.edu}\affiliation{Southern University, 
Baton Rouge, LA 70813}
\author{R.M. Gunasingha}\altaffiliation{Present address: Box 3155, Duke University 
Medical Center, Durham, NC 27710}\affiliation{Southern University, Baton Rouge, LA 70813}
\author{R.L. Imlay}\affiliation{Southern University, Baton Rouge, LA 70813}
\author{K.D. Muhammad}\affiliation{Southern University, Baton Rouge, LA 70813}
\author{S.V. Ter-Antonyan}\affiliation{Southern University, Baton Rouge, LA 70813}
\author{X. Xu}\affiliation{Southern University, Baton Rouge, LA 70813}
%Collaboration name if desired (requires use of superscriptaddress
%option in \documentclass). \noaffiliation is required (may also be
%used with the \author command).
%\collaboration can be followed by \email, \homepage, \thanks as well.
%\collaboration{}
%\noaffiliation

\date{\today}

%\linenumbers
\begin{abstract}
% insert abstract here
We investigate the discovery potential of cubic kilometer
neutrino observatories such as IceCube to set stringent limits on the forbidden 
decays
$\eta\to\nu_e\bar\nu_e$ and $\eta\to\nu_{\tau}\bar\nu_{\tau}$.
The signatures for these decays are cascade events
resulting from the charged-current reactions of
$\nu_e, \nu_{\tau}$, $\bar\nu_e$ and $\bar\nu_{\tau}$ on nuclei in such
detectors.
Background cascade events are mainly due to $\nu_e$'s
from atmospheric $\mu$, $K^+$, and $K^{0}_S$ decays and to a lesser extent
from atmospheric $\nu_{\mu}$
neutral current interactions with nuclei.
A direct upper limit for the branching ratio
$\eta \to \nu_{e,\tau} \bar\nu_{e,\tau}$ of
$6.1 \times10^{-4}$ at $90\%$ CL can be achieved.

\end{abstract}

% insert suggested PACS numbers in braces on next line
% insert suggested keywords - APS authors don't need to do this
%\keywords{}

%\maketitle must follow title, authors, abstract, \pacs, and \keywords
\maketitle

% body of paper here - Use proper section commands
% References should be done using the \cite, \ref, and \label commands
\section{Introduction}
The observation of the decay $\eta\to\nu\bar\nu$ or $\pi^0\to\nu\bar\nu$
would imply new and interesting physics. The $\eta$ as well as the $\pi^0$ have
zero spin and odd intrinsic parity, i.e. $J^P=0^-$, and thus conservation of
momentum and angular momentum
require that the decay products $\nu$ and $\bar\nu$ possess
the same helicity. This decay provides an ideal laboratory to search for
the pseudoscalar (P)
weak interaction, because only the P interaction allows the selection rule for
the $0^+\to 0^-$ transition for nearly massless neutrinos and
antineutrinos. Other exotic effects such as the presence of a right-handed
weak current through the exchange of a $Z^{0}_R$
would also allow such decays.
The information derived from $\pi^0\to \nu\bar{\nu}$ and
$\eta\to \nu\bar{\nu}$ are complementary because the former is
sensitive only to the isovector neutral-current (NC) interactions while
the latter
is sensitive to the isoscalar NC interactions \cite{peter}.
Furthermore, $\pi^0$ decays involve only u- and d-quarks while $\eta$ decays
additionally involve the s-quark and perhaps other heavier quarks.
If the $Z^0$ couples to a massive neutrino with the standard
weak-interaction strength, the branching ratio (BR) for
$\pi^0\to\nu_\tau \bar\nu_\tau$ and $\eta\to\nu_\tau \bar\nu_\tau$ have
maximum values of
$5.0\times10^{-10}$ and $1.3\times10^{-11}$ \cite{arn}, respectively at
the $\nu_\tau$ mass upper limit of
$m_{\nu_\tau} = 18.2$ MeV/$c^2$ \cite{pdg}.
It is noteworthy that BRs of $\approx 2 \times 10^{-18}$ and
$\approx 2 \times 10^{-15}$ are allowed
within the Standard Model (SM) for $\pi^0 \to \nu\bar\nu\gamma$ and
$\eta \to \nu\bar\nu\gamma$, respectively \cite{arn}.
\section{Existing limits}
To date no exclusive limits have been set on $\eta \to \nu\bar\nu$ in any
experiment. The Particle Data Group (PDG)\cite{pdg} reports an inclusive
upper limit of
$\Gamma(\eta\to invisible)/\Gamma(\eta\to \gamma\gamma)
< 1.65 \times 10^{-3}$ from the BES-II Collaboration\cite{bes},
corresponding to an upper limit on the BR for $\eta\to invisible$ of
$6.0 \times 10^{-4}$.
The BES-II results are inclusive results obtained by
using $58 \times 10^6$ $J/\psi \to \phi\eta$ decays.

Possible $\eta\to invisible$ decay products
could be Light Dark Matter (LDM) particles or light neutralinos.
These LDM particles
may have an adequate relic density to account for the non-baryonic mass
of the universe. Our estimated IceCube limits will be
complementary to BES-II limits since the SM neutrinos would be a component
of BES-II reported inclusive measurements.
Limits on $\pi^0\to\nu_\alpha\bar{\nu_\alpha}$
($\alpha = \nu_e, \nu_\mu, \nu_\tau$) are more common.
An experimental upper limit, $\Gamma(\pi^0\to\nu_e\bar{\nu_e})/
\Gamma(\pi^0\to all) < 1.7 \times 10^{-6}$  at $90\%$ confidence
level (CL), was set by Dorenbosch et al. \cite{dorn}. 
The LSND Collaboration \cite{lsnd} has set an upper limit on the BR for 
$\pi^0\to\nu_{\mu}\bar{\nu_{\mu}}$ of $1.6 \times 10^{-6}$ at $90\%$
CL. In the tau neutrino channel, Hoffman has set a limit of
$\Gamma(\pi^0\to\nu_{\tau}\bar{\nu_{\tau}})/
\Gamma(\pi^0\to all) < 2.1 \times 10^{-6}$ at $90\%$ CL \cite{hoff}.
An inclusive search for $\pi^0 \to \nu\bar\nu$ using
$K^+ \to \pi^+\pi^0$ has set an upper limit of $2.7 \times 10^{-7}$
at $90\%$ CL \cite{arto}, (see PDG for detail).

\section{Calculations for $\eta\to\nu_{e,{\tau}}\bar\nu_{e,{\tau}}$}
Because of their enormous mass, cubic kilometer neutrino detectors such as
IceCube, offer a new opportunity to search for such exotic decays with
competitive results compared to those obtained from accelerator-based
experiments. IceCube, presently near completion 
at the South Pole, will contain 4800 digital optical modules (DOM) mounted on
80, 1-km strings. The active target consists
of approximately $4.2 \times 10^{37}$ $^{16}O$ atoms and $8.4 \times 10^{37}$
H atoms.
We have performed calculations using the
CORSIKA Extensive Air Shower simulation code, version 6.72, to estimate the
number of $\eta$ mesons produced in the
atmosphere \cite{CORSIKA}.
The primary nucleon energy spectrum was calculated based on a sum of
the power law approximations for the elemental primary
energy spectra,
\begin{equation}
I(E)=\sum A_i\Phi_{A_i}(EA_i)^{-\gamma_A}.
\end{equation}
The parameters $\Phi_{A_i}$ and $\gamma_A$ for
$i=1,\dots 28$ primary nuclei with mass number $A_i$
were obtained from corresponding approximations
of balloon and satellite data \cite{Wiebel}. The resulting 
nucleon energy spectrum of expression (1) can then be written as;
\begin{equation}
I(E)=(0.110\pm0.006)E^{-2.74\pm0.02}
\end{equation}
in units of $(m^2\cdot s\cdot sr\cdot TeV)^{-1}$. The simulation program
was tested by comparing the
simulated neutrino and anti-neutrino energy spectra for two zenith angles
($\theta=0$ and $\theta=60^0$) with the corresponding spectra
of Gaisser and Honda \cite{gais}.
\begin{figure}
\includegraphics[width=8cm,height=8cm]{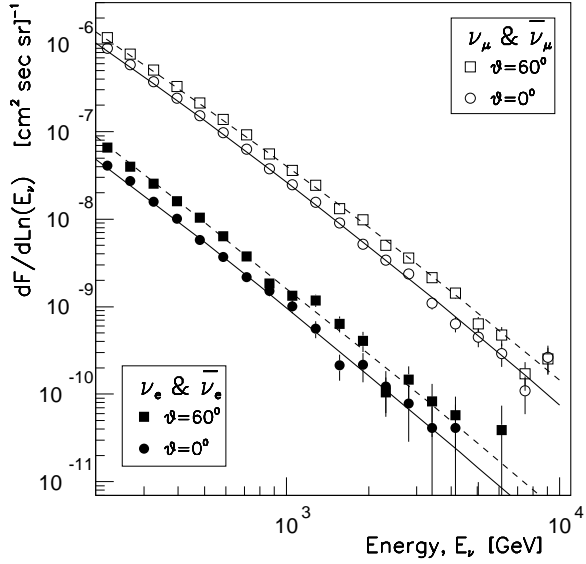}%---------------------- fig. 1
\caption{Atmospheric $\nu_e(\bar\nu_e)$, and $\nu_{\mu}(\bar\nu_{\mu})$
energy spectra for two zenith angles compared with those of Gaisser and
Honda \cite{gais} shown with solid and dashed lines.\label{fig:test}}
\end{figure}

Shown in Fig. \ref{fig:test} are the sum of fluxes of
$\nu_{\mu}$ and $\bar\nu_{\mu}$, above 200
GeV at zero degree (open circles) and at $60^\circ$ (open squares). Also
shown are the sum of fluxes of $\nu_e$ and $\bar\nu_e$ above 200 GeV
at zero degree (solid circles) and at $60^\circ$ (solid squares).
These calculations, as shown in Fig.\ref{fig:test}, agree well with those of
Gaisser and Honda \cite{gais}.
\begin{figure} %--------------------------fig.2
\includegraphics[width=8cm,height=8cm]{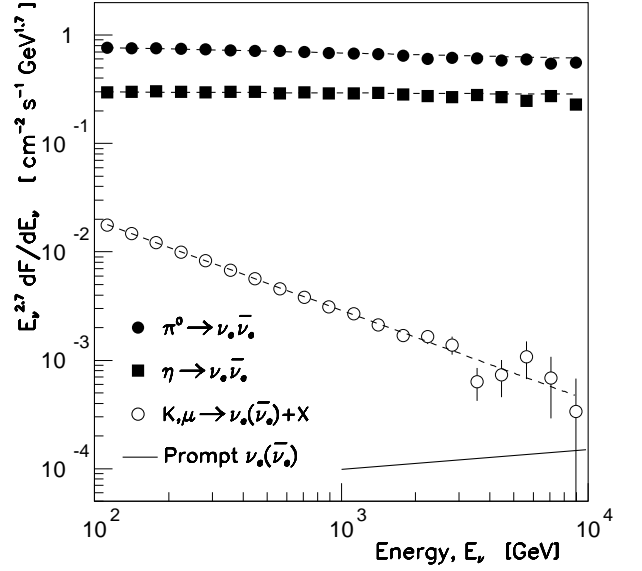}
\caption{The expected energy spectrum of $\nu_e(\bar\nu_e)$
from decays of atmospheric $K$ and $\mu$ is shown
by open circles. The solid line shows the contribution of $\nu_e(\bar\nu_e)$ from
charm decays\cite{reno}. 
The $\nu_e(\bar\nu_e)$ energy spectra from possible decay
modes $\pi^0 \to \nu_e(\bar\nu_e)$ (solid circles) and $\eta \to \nu_e(\bar\nu_e)$ 
(solid squares) are shown assuming a $100\%$ BR.}
\label{dide}
\end{figure}
\begin{figure} %--------------------------fig.3
\includegraphics[width=8cm,height=8cm]{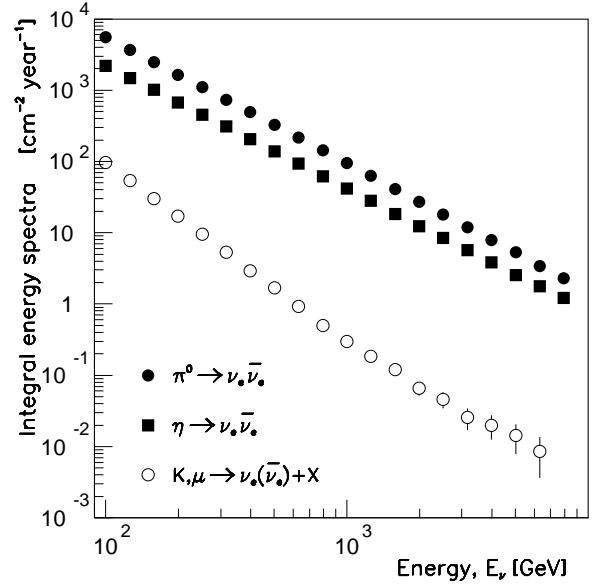}
\caption{$\nu_e$ integral energy spectra (see Fig~2)
in the units of ($cm^2\cdot year$)$^{-1}$.}
\label{i_sp}
\end{figure}
The $\nu_\mu(\bar\nu_\mu)$ rate in Fig.\ref{fig:test} is an order of magnitude larger than
$\nu_e(\bar\nu_e)$ rate because pions and kaons decay mostly to $\nu_\mu(\bar\nu_\mu)$
and not to $\nu_e(\bar\nu_e)$. The largest sources of $\nu_e(\bar\nu_e)$ are
$K^{\pm}_{e3}$ decay, i.e. $K^\pm \to \pi^0 e^\mp \nu_e$, $BR = 5.08\%$, and
$K^0_{e3}$ decay, i.e. $K^0_{L} \to \pi^\pm e^\mp \nu_e$, $BR = 40.55\%$ and
to a lesser extent $\mu^\pm$ decay. Due to its long lifetime most $\mu^\pm$
reach the ground before decaying. Because the atmospheric
$\nu_e(\bar\nu_e)$ background is significantly less than the atmospheric
$\nu_\mu(\bar\nu_\mu)$ background, we concentrate on searches for
$\eta \to \nu_e(\bar\nu_e)$ and $\eta \to \nu_\tau(\bar\nu_\tau)$.
Interactions of high energy atmospheric neutrinos in IceCube can be
classified as
events with a long muon track or as cascade events with very localized energy
deposited in the detector.
The muon events are due to charged-current (CC)
$\nu_\mu$ interaction while the cascade events are mainly from CC $\nu_e$
and to a lesser extent from NC $\nu_e$ and $\nu_\mu$.
Neutrino absorption in the Earth was also taken into account
using the neutrino mean free path
$\lambda_{\nu}=1/(N_A\rho(\theta)\sigma(E_{\nu}))$,
where $N_A$ is the Avogadro's number, $\rho(\theta)$ is the average density of
the Earth in g/cm$^3$ \cite{and} for a neutrino traversing the Earth at angle 
$\theta$ and $\sigma $ is the $\nu$-nucleon cross section at neutrino energy $E_{\nu}$
using a parton distribution functions from CTEQ6 \cite{mary}.

In Fig. \ref{dide} the expected sum of the
$\nu_e$ and $\bar\nu_e$ energy
spectrum from the standard K and $\mu$ decay modes is presented
(open circles). 
Fig. \ref{dide} also shows the prompt $\nu_e(\bar\nu_e)$s from 
charm decay. 
This contribution becomes significant by 10 TeV and dominates at high 
energies\cite{reno}.
Also shown are neutrino energy spectra for the possible
decay modes $\pi^0\to\nu_e\bar{\nu_e}$ (solid circles) and
$\eta\to\nu_e\bar{\nu_e}$ (solid squares), assuming $100\%$ BR.
The energy spectra of neutrinos
from a $\pi^0$ and $\eta$ SM-forbidden decays have nearly the same shape
as the primary nucleon spectrum ($\gamma\simeq-2.7$) whereas the energy
spectrum of neutrinos from $K$ and $\mu$ decays is significantly steeper
($\gamma\simeq-3.58$). This is because both $\pi^0$ and $\eta$ 
with very short lifetimes ($\tau_{\pi^0} = 8.4 \times 10^{-17} s$ and
$\tau_{\eta} \approx 5.0 \times 10^{-19} s$) do not interact and
lose energy before decaying, while charged pions and kaons with much longer
lifetimes interact
substantially with the atmosphere before decaying.  
The corresponding integral spectra are shown in Fig. \ref{i_sp}.
The flux of background neutrinos from the lower hemisphere
($\cos{\theta<0}$) is
practically equal to the flux from the upper hemisphere because
neutrino absorption in the Earth is approximately
compensated by the larger atmosphere depth in the case of upward-going
neutrinos originating from the northern hemisphere.

The flux of neutrinos induced by $\pi^0$ and $\eta$ decays
slightly depend on atmospheric depth in the range of 700-1000 g/cm$^2$.
The CORSIKA generated
neutrinos from $\pi^0$ and $\eta$ decays with a CTEQ6
parton distribution model and the corresponding
cross sections for CC and NC were calculated according to the formalism
employed by Reno \cite{mary}. The expected rate of detectable $\nu_e$ events
for the IceCube detector was then calculated using a GCALOR simulation MC
program \cite{gab}.
\begin{figure}   %--------------------------fig. 4
\includegraphics[width=8cm,height=8cm]{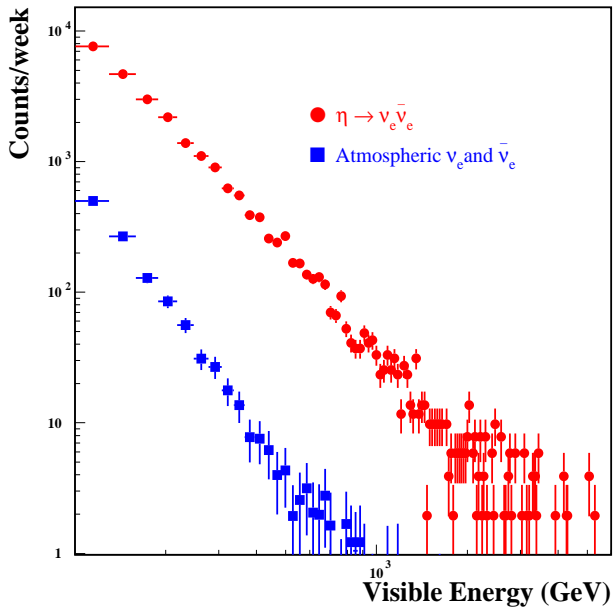}
\caption{GEANT output for the spectrum of Fig. 3. The trigger 
requirement of at least 8 PMTs has been applied.\label{etasp}}
\end{figure}
\begin{figure} %--------------------------fig. 5
\includegraphics[width=8cm,height=8cm]{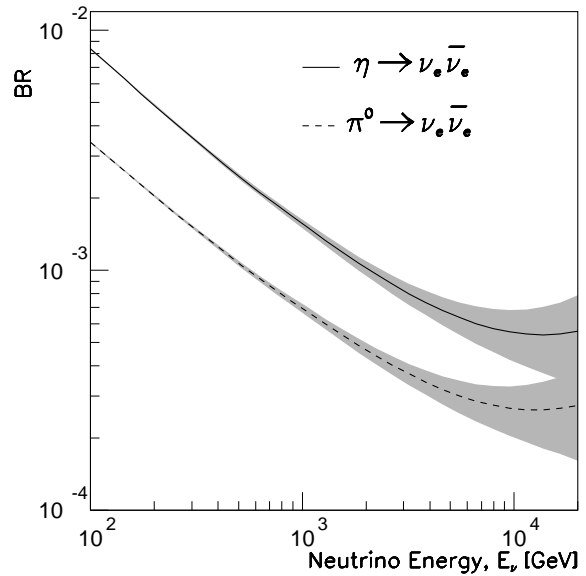}
\caption{Measurable upper limits (lines) for the BR of 
$\pi^0,\eta\to\nu_e\bar\nu_e$ decays versus neutrino energy for 5 years 
operating live time of the IceCube detector. The shaded areas show the 
statistical uncertainties.} 
\label{bree.eps}
\end{figure}

\section{GCALOR Calculations and Limits}
The 3-momenta of these events for $e$ and $\tau$ leptons were written to a file
which was then read by GEANT and electron and
$\tau$ transport with their accompanying hadron, a proton in this case,
were simulated. The interaction vertices were distributed uniformly
throughout the detector volume. The GEANT Cerenkov code together with
the input IceCube geometry with average PMT quantum efficiency
as well as an ice model with appropriate absorption and scattering
\cite{wosch} simulated the
hit PMTs. The resulting trigger efficiencies for number of hit PMTs $\geq8$
are 0.75 and 0.70 for $\eta \to \nu_{e,\tau}\bar\nu_{e,\tau}$ and the
atmospheric
background due to K and $\mu$ decays, respectively. Fig. \ref{etasp} shows
the energy distributions of the possible signal and the background events. We
also estimated the atmospheric background due to cascades from $\nu_{\mu}$
NC interaction with nuclei as well as CC $\nu_{\mu}$
interactions where the muon is not detected due to edge effects. These types of
events have a trigger efficiency of 0.30 and contribute only at a $15\%$ level
and have been included in the atmospheric neutrino background. The atmospheric
$\nu_{\tau}$ contribution to the background are small below 10 TeV \cite{reno}.
Furthermore, above a few TeV prompt $\nu_e(\bar\nu_e$'s from charm decay must 
be taken into consideration\cite{reno}.
The amount of MC data shown corresponds to one week of data taking with an
80-string IceCube detector configuration.
Fig. \ref{bree.eps} shows the expected upper limits on BRs  
for $\pi^0\to\nu_e\bar{\nu_e}$ and
$\eta\to\nu_e\bar{\nu_e}$ decays that could be obtained from
5 years of measurements by the IceCube detector.
The computations were
performed at a $90\%$ CL using the expression,
%\begin{equation}
%BR \le \frac{\sqrt{{I_{\nu}(K,\pi,\mu\to\nu(\bar{\nu})+X)}
%+(\Delta_{sys})^2}}{I_{\nu}(\pi^0, \eta\to\nu\bar{\nu})}
%\end{equation}
\begin{equation}
BR \le \frac{\sqrt{I_{Bkgr}
+(\Delta_{sys})^2}}{I_{\nu}(\pi^0, \eta\to\nu\bar{\nu})}
\end{equation}

Where $I_{Bkgr}$ is the number of background 
events from K, $\pi$, $\mu$ and prompt charm decays. $\Delta_{sys}$ is the 
systematic uncertainty in the number of background 
events, 
and $I_{\nu}(\pi^0, \eta\to\nu\bar{\nu})$ is the estimated number of $\eta$ 
decay events assuming $100\%$ BR. 
Note in the above expression and the figure, the systematic uncertainties
are mainly from two sources, the primary cosmic ray flux and cross sections. 
These
uncertainties are energy dependent and a full analysis of them would include
uncertainties in flux calculations and uncertainties associated with 
particle production cross sections. We estimated these uncertainties based
on the energy-dependent flux uncertainties reported by Agrawal, et al. \cite{agra}
and those reported by Derome \cite{derom}.
Uncertainties for the neutrino production cross sections reported by the two 
references is $15\%$.
The relative primary nucleon flux systematic error is due to uncertainties
in Equation 2 and 
$I_{Bkgr}$ simulations and can be approximated by
\begin{equation}
\frac{\Delta I}{I}=\sqrt{0.1^2 + \Big(0.02 \ln(\frac{E_{\nu}k}{1000GeV})\Big)^2},
\end{equation}
where $\ln k=<\ln(E/E_{\nu})>\simeq2.75\pm0.05$.  
Table \ref{tab:uncer} shows these uncertainties for the primary flux and the 
neutrino production cross sections using the reported values of reference 
\cite{agra}. 
\begin{table}
\caption{Uncertainties associated with primary flux and neutrino production 
cross sections.}
\begin{tabular}{c c c c}
\hline
  $E_{\nu}$ (GeV)   & $10^2$ & $10^3$ & $10^4$\\
\hline
     Primary Flux   &  0.10   & 0.11   &  0.14\\
     $\sigma_{\nu}$ &  0.15   & 0.15   &  0.15\\
     Overall        &  0.18   & 0.19   &  0.21\\
\hline
\end{tabular}
\label{tab:uncer}
\end{table}
 
Uncertainties in table \ref{tab:uncer} are the most conservative estimates
that contribute to the limits on $\eta\to\nu_{e,\tau}\bar{\nu_{e,\tau}}$. As shown in Fig. \ref{bree.eps}, the most stringent limits are obtained from 
neutrinos above $\approx$ 9 TeV. 
\section{Summary}
In summary, we have investigated the IceCube discovery potential
for setting stringent limits
on the $\eta\to \nu_e\bar\nu_e$ and $\eta\to \nu_{\tau}\bar\nu_{\tau}$.
Our studies show that a direct upper
limit of $6.1 \times 10^{-4}$ at $90\%$ CL at neutrino energies above 9 TeV 
for both $\eta$ decay to two $e$ neutrinos
or two $\tau$
neutrinos can be obtained. This
limit is complementary to the limit set by the inclusive $\eta\to nothing$
measurements of reference 4.
\begin{acknowledgments}
The authors gratefully acknowledge a MRE grant from the National Science
Foundation through a subcontract from the University of Wisconsin Board of
Regents under the contract No. G067933. We are also grateful for valuable
comments by Professor Francis Halzen and Dr. William C. Louis, III.
\end{acknowledgments}

\end{document}